\documentclass{optica-article}

\journal{opticajournal} 

\articletype{Research Article}

\usepackage{lineno}
\usepackage{comment}

\begin{document}

\title{Widefield pump-probe microscopy with coherent background subtraction by angle-compensated temporal shearing}

\author{Matthew Sheinman,\authormark{1,*} 
Mark Polkovnikov, \authormark{1}
Luke Saunders, \authormark{1} 
Pajo Vujkovic-Cvijin, \authormark{2}
Shyamsunder Erramilli,\authormark{1,2} 
Lawrence D. Ziegler,\authormark{2,3} 
Koustav Kundu, \authormark{2,3}
Ramprasath Rajagopal, \authormark{1}
Bingying Zhao,\authormark{4} 
Christopher McMahon,\authormark{1}
Mi K. Hong,\authormark{1}
and Jerome Mertz\authormark{2,5}}

\address{\authormark{1}Department of Physics, Boston University, 590 Commonwealth Avenue, Boston, Massachusetts 02215, USA\\
\authormark{2}Photonics Center, Boston University, 8 St. Mary’s St., Boston, Massachusetts 02215, USA\\
\authormark{3}Department of Chemistry, Boston University, 590 Commonwealth Avenue, Boston, Massachusetts 02215, USA\\
\authormark{4}Department of Electrical and Computer Engineering, Boston University, 44 Cummington Mall, Boston,MA 02215, USA\\
\authormark{5}Department of Biomedical Engineering, Boston University, 44 Cummington Mall, Boston, Massachusetts 02215, USA}

\email{\authormark{*}sheinman@bu.edu} 


\begin{abstract*} 

Pump-probe microscopy enables label-free imaging of structural and chemical features of samples. However, signals in pump-probe microscopy are typically small and often must be measured in the presence of large backgrounds. As a result, achieving measurements with a high signal-to-noise ratio is challenging, particularly when using sensors that are easily saturated, such as CMOS cameras. We present a method for enhancing signal-to-noise ratio while avoiding detector saturation. In this approach, temporally separated (sheared) reference and probe pulses transmit through a sample before and after the arrival of a pump pulse. The probe and reference pulses are then temporally recombined with opposing phases and nearly matched amplitudes, resulting in interferometric background subtraction. This recombining operation is performed by a novel common-path interferometer. Unlike previous techniques for temporal shearing, this interferometer demonstrates negligible phase and group delay dispersion with angle of incidence, allowing convenient widefield imaging. To our knowledge, this is the first common-path interferometer with such a property. We demonstrate the technique by measuring transient absorption signals in gold nanorod films with a signal-to-background ratio enhanced by over $100$\% and a signal-to-noise ratio enhanced by about $70$\%.

\end{abstract*}

\section{Introduction}

Pump-probe microscopy measures modulations in a probe beam following interaction with a sample excited by a pump beam \cite{Fischer2016,Hu2019,Zhu2020,dong2017pump}, enabling the study of excited-state dynamics without the introduction of exogenous labels. Representative techniques include coherent Raman scattering (CRS) \cite{Zhang2014,Hu2019}, photothermal (PT) \cite{Zhang2016}, and transient absorption (TA) \cite{Virgili2012,Li2015,Chen2018,Dong2019,Zhu2020,Huang2018} microscopy. Striking applications have included monitoring drug uptake in diseased cells with CRS \cite{Tipping2016}, PT imaging of protein and lipid distributions in living organisms \cite{Zhang2016,zong2021background}, and the study of melanoma with TA \cite{Matthews2011,Tyler2015}. Typically, signals in pump-probe microscopy are small and must be measured in the presence of a large background. A key challenge is to increase spatial and temporal imaging resolution while maintaining the capacity to detect such small signals.

Approaches to pump-probe microscopy can be divided into two broad categories \cite{Zhu2020}. In the scanning geometry, focused pump and probe beams with high repetition rates and low pulse energies are swept over a sample. Measurements are then recorded serially with a single element detector. In contrast, in the widefield geometry, unfocused pump and probe beams with high pulse energies excites and probe the entire sample simultaneously \cite{Fantuzzi2023,Bai2019,Matthews2017,zong2021background,Massaro2016}. The ensuing modulation is then detected with an array detector, such as a conventional CMOS camera. The widefield scheme avoids the mechanical complexity of a high-speed scanner and enables higher spatiotemporal sampling rates. Despite this, the scanning geometry has remained dominant, largely because the reported sensitivity of the widefield geometry has been comparatively small.

Two obvious sources of noise contribute to this reduced sensitivity. First, the reduced repetition rate in a widefield geometry results in increased shot-to-shot power fluctuations, referred to as relative intensity noise (RIN). This is in accordance with the well known empirical trend that RIN scales inversely with the laser repetition rate. While it is possible to mitigate the influence of this noise with balanced detection, this remains challenging when the RIN is spatially varying \cite{Hormann2024}. More fundamentally, sensitivity in the widefield geometry is limited by shot noise. It is well known that the signal-to-noise ratio (SNR) for shot-noise-limited measurements scales as $1/\sqrt{N}$, where $N$ is the number of collected photoelectrons. Current CMOS cameras typically feature full well capacities on the order of $10^4$ electrons, corresponding to a modest sensitivity of $10^{-2}$. Even state-of-the-art prototypes feature full well capacities of only $10^6$ electrons \cite{Yaqoob2016}. Increasing sensitivity then requires averaging counts from multiple pixel read-outs in either space or time, which comes at the expense of spatial and temporal resolution.

We demonstrate the possibility of enhancing sensitivity without sacrificing resolution using a technique of temporal interferometry to reduce background. The technique entails shearing a pulse into reference and probe pulses that arrive before and after the pump pulse, and then recombining (unshearing) these with a $180^\circ$ phase shift and nearly matched amplitudes, resulting in background reduction by coherent background subtraction (CBS). This CBS prevents detector saturation even while allowing the light power incident on the sample to be increased, thus increasing SNR. As we will show, in the ideal case, the SNR is ultimately limited by the number of photons incident on the sample.

For samples whose excited states exhibit short lifetimes ($\lesssim 10$ ps), the temporal shearing and unshearing operations can be performed conveniently with birefringent crystals comprising a common-path interferometer. This approach has been previously applied to transient absorption microscopy of gold nanoparticles \cite{Dijk2005,Orrit2007}; however, it was noted that the phase shift between probe and reference beams depended on the angular directions of the wave vectors $\mathbf{k}_{i}$ incident on the crystals. To avoid this problem of angular dispersion, the sample had to be scanned through fixed, focused probe and reference beams, such that the beams could be collimated at the crystals. Such a collimation constraint  precluded the possibility of widefield imaging.

Applications of birefringent crystals to microscopy for other purposes have inevitably encountered similar difficulties with angle dependent phase shifts. Instead of reducing background with a phase shift near $180^\circ$, efforts have been made to perform ultrafast phase-contrast microscopy in a scanning geometry using a phase shift near $90^\circ$ \cite{Coleal2024,Smith2024}. This has enabled, for instance, low frequency impulsive stimulated Raman scattering microscopy with enhanced sensitivity. However, it was noted that the nominal  $90^\circ$ phase bias varied with scan angle. In addition to pump-probe microscopy, birefringent interferometers have been used for widefield Fourier Transform spectroscopy \cite{Candeo2019,Perri2019}. Here, too, it was noted that imaging performance was degraded by the angular dispersion in $\mathbf{k}_{i}$, and that this angular dispersion had a hyperbolic profile.

Flattening the angular dispersion in $\mathbf{k}$ is a key requirement for achieving high-resolution widefield temporal interferometry over large fields of view (FOVs). To this end, we introduce a method of angle-compensated temporal shearing (ACTS) to perform such flattening. We derive a theoretical model predicting the observed angular dispersion relation for uniaxial birefringent crystals. We then apply this model to the design of a common-path ACTS interferometer that is angular-dispersion-free in principle, and in practice reduces angular dispersion significantly. When applied to transient absorption microscopy with CBS, our ACTS interferometer is shown to enable enhancements of SBR by over $100$\% and SNR by about $70$\%, limited only by the extinction ratio of our polarizing beamsplitting optics.

\section{Methods}

\subsection{Principle of Coherent Background Subtraction}

\begin{figure}[h]
    \label{CBS_theory_figure}
    \includegraphics[width=13.2cm]{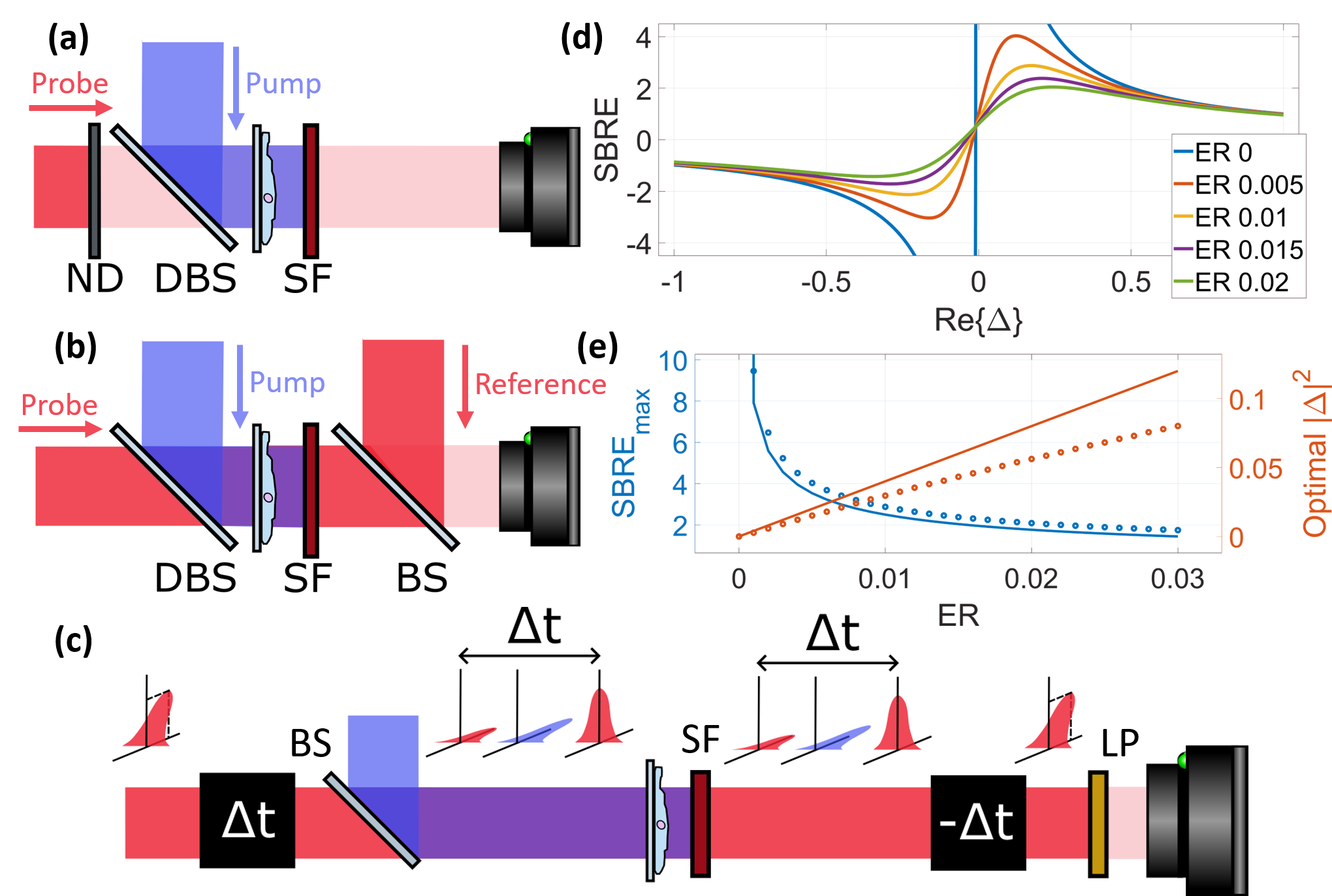}
    \caption{(a) Conventional widefield pump-probe imaging, with imaging optics omitted for simplicity. To avoid saturating the camera, probe photons must be discarded with e.g. a neutral density filter (ND). DBS dichroic beamsplitter; SF spectral filter. (b) Coherent background subtraction. Saturation is avoided by reducing intensity interferometrically with a reference beam, introduced in this thought experiment with a beamsplitter (BS). (c) Coherent background subtraction separating cross-polarized probe and reference beams in time instead of space. LP linear polarizer. (d) The expected $SBRE$ as a function of $\Delta$ for different values of ER, simulated with  $t = 0.01 - 0.01i$, $\epsilon_{S} = \frac{1}{2}\epsilon_{BG}$, and $\mathrm{Im}\{\Delta\} = 0$. (e) Behavior of the $SBRE$ peak as a function of  ER. Dotted lines are simulated from the same parameters in Fig. 1d. Solid lines are approximations obtained from Eqs. \ref{eq12} and \ref{eq13}.}
\end{figure}

To understand the benefits of CBS, we first consider the case of conventional widefield pump-probe imaging [Fig. 1a]. When operating close to detection saturation, the field and intensity at the camera are given by
\begin{equation} \label{eq1}
    \begin{split}
        E_{cam} & = E_{sat}(1 + t) \\
        I_{cam} & = I_{sat}(1 + 2\mathrm{Re}\{t\} + |t|^{2})
    \end{split}
\end{equation}
where $t$ corresponds to the perturbation in complex amplitude transmittance of the sample induced by the pump beam, in general quite small compared to the background transmittance (assumed here to be $1$ for simplicity).

Measurement noise in this case is determined by the saturation intensity $I_{sat}$ (combination of RIN and shot noise), along with additional noise depending on transient dynamics; we denote the standard deviation of noise as $\sigma$. The SBR and SNR in the conventional case are then
\begin{equation} \label{eq2}
    \begin{split}
        SBR_{conv} & \approx 2\mathrm{Re}\{t\} \\
        SNR_{conv} & \propto SBR_{conv} \frac{I_{sat}}{\sigma}
    \end{split}
\end{equation}
where higher order terms in $t$ are assumed negligible. SNR is evidently limited here by the small values of $\mathrm{Re}\{t\}$ and $I_{sat}$, and the only recourse for increasing SNR is to increase the number of photoelectrons per measurement by averaging in space or time, which degrades spatiotemporal resolution.

We next consider achieving the same $I_{sat}$ at the camera, but this time with the introduction of CBS [Fig. 1b]. In order to illustrate this approach, we consider the introduction of a spatially separated reference beam. In our actual experiment, described below, these fields will be separated in time. The probe and reference fields $E_{probe}$ and $E_{ref}$ combine coherently, obtaining
\begin{equation} \label{eq3}
    \begin{split}
        E_{cam} & = E_{probe}(1 + t) - E_{ref}= E_{probe}(\Delta + t)  \\
        I_{cam} & = I_{probe}(|\Delta|^{2}
        + 2\mathrm{Re}\{\Delta\}\mathrm{Re}\{t\}
        + 2\mathrm{Im}\{\Delta\}\mathrm{Im}\{t\}
        + |t|^{2})
    \end{split}
\end{equation}
where we have introduced the parameter $\Delta = 
\frac{E_{probe} - E_{ref}}{E_{probe}}$, which is complex in general. Note that CBS allows the possibility of phase sensitive measurements when $\Delta$ has a nonzero imaginary part, which can be achieved by inducing a phase shift between probe and reference fields.

If $\Delta$ is set such that $|\Delta|^{2}I_{probe} \approx I_{sat}$, the noise $\sigma_{st}$ and background remain the same as in the conventional case, leading to
\begin{equation} \label{eq4}
    \begin{split}
        SBR_{CBS} & \approx 2\frac{\mathrm{Re}\{\Delta\}\mathrm{Re}\{t\}
        + \mathrm{Im}\{\Delta\}\mathrm{Im}\{t\}}
        {|\Delta|^{2}} \\
        SNR_{CBS} & \propto SBR_{CBS} \frac{I_{sat}}{\sigma}
    \end{split}
\end{equation}

We note that SBR now grows as $1/|\Delta|$. The SBR enhancement (SBRE) and SNR enhancement (SNRE) are obtained from the ratios of Eqs. \ref{eq4} and \ref{eq2}. In particular, for real $\Delta$ we have
\begin{equation} \label{eq5}
    SBRE = SNRE = \frac{1}{|\Delta|}
\end{equation}

Since $|\Delta|^{2} \approx I_{cam}/I_{probe}$, the term $1/|\Delta|$ can be interpreted as $1/\sqrt{T_{int}}$, where $T_{int}$ is the interferometer transmittance. For example, if the interferometer transmits 1\% of incident probe power, this simple model predicts a SBRE and SNRE of $10$. Note that the above assumes $\sigma$ remains unchanged when using CBS; in practice, this assumption may not hold. Generally, then, the SBRE represents an upper limit for the SNRE.

In our case, we perform CBS by a method of temporal shearing [Fig. 1c]. Cross-polarized probe and reference pulses are separated by a time $\Delta t$  before arriving at the sample, such that the probe pulse is modulated by pump-induced changes in the sample, whereas the reference pulse is not. After transmission through the sample, probe and reference pulses are recombined in time and add coherently; Eqs. \ref{eq1}-\ref{eq5} hold as before. Note that both probe and reference beams experience identical modulation by static features. This allows a high degree of interferometric visibility even when samples exhibit spatially complex static transmissions, a feature not shared by microscopes suppressing background through conventional darkfield techniques \cite{zong2021background}.

If the dominant source of noise is shot noise, then $\sigma \propto \sqrt{I_{sat}}$, and, for real $\Delta$, Eqs. 4 and 5 can be used to show that $SNR_{CBS} \propto 2\mathrm{Re}\{t\}\sqrt{I_{probe}}$, independent of $I_{sat}$. This is the same scaling that would be anticipated for a shot-noise limited experiment which was not constrained by a sensor's full well capacity. In the ideal case, then, CBS removes the limit on SNR imposed by equipment, leaving only the fundamental limit imposed by Poisson statistics. It is worth exploring the degree to which this limiting case can be demonstrated.

\subsection{Practical Coherent Background Subtraction}

Eq. \ref{eq5} suggests unbounded performance improvements as $\Delta$ tends to zero.  In practice, SBRE and SNRE are limited by the maximum extinction achievable with the interferometer. Intuitively, if some fraction of the camera's readout is due to fields that failed to add coherently, we should expect that portion to not benefit from CBS. To estimate the effect of imperfect extinction, we include incoherent contributions to the beams, obtaining
\begin{equation}
    \begin{split}
        E_{cam} & = E_{probe}(\Delta + t)
        \oplus \sqrt{I'_{probe}}(1 + t) \oplus \sqrt{I'_{ref}} \\
        I_{cam} & = I_{probe}(|\Delta|^{2} + 2\mathrm{Re}\{\Delta\}
        \mathrm{Re}\{t\} + 2\mathrm{Im}\{\Delta\}
        \mathrm{Im}\{t\} + |t|^{2}) \\
        & + I'_{probe}(1 + 2\mathrm{Re}\{t\} + |t|^{2}) + I'_{ref}
    \end{split}
\end{equation}
where $I'_{probe}$ and $I'_{ref}$ represent the incoherent parts of the probe and reference intensities, and the symbol $\oplus$ represents incoherent addition.

Note that some of the terms associated with the incoherent contributions are modulated by $t$ (producing signal) while others are not (producing background). The SBR can thus be expressed as

\begin{equation} \label{eq9}
    SBR_{CBS} = \frac{2(\mathrm{Re}\{\Delta\} + \epsilon_{S})\mathrm{Re}\{t\}
     + 2\mathrm{Im}\{\Delta\}\mathrm{Im}\{t\} + (1 + \epsilon_{S})|t|^{2}}
    {|\Delta|^{2} + \epsilon_{BG}}
\end{equation}
where $\epsilon_{S} = I'_{probe}/I_{probe}$ represents a small incoherent component modulated by the signal, and $\epsilon_{BG} = (I'_{probe} + I'_{ref})/I_{probe}$ represents a small incoherent component which is not as modulated, and thus contributes to background.  In a real experiment, the method used to tune $\Delta$ may well also alter the values of  $\epsilon_{S}$ and $\epsilon_{BG}$. However, for purposes of deriving reasonable estimates, we assume they are fixed.

SBRE can be found as before from the ratio of the predicted SBR's with and without CBS. By inspection, it is clear that the term $\epsilon_{S}$ in the numerator of Eq. \ref{eq9}  causes only slight deviations from the trend predicted by Eq. \ref{eq5}. On the other hand, even small values of $\epsilon_{BG}$ in the denominator can cause substantial reductions in $SBR_{CBS}$. To express $\epsilon_{BG}$ in terms of experimentally measurable parameters, we consider the ratio of measured intensities when the probe and reference fields are in and out of phase, with the pump beam blocked (i.e., $t = 0$). Calling this the extinction ratio (ER), it follows that
\begin{equation} \label{eq10}
    \epsilon_{BG} = 4 \, \text{ER}
\end{equation}

Fig. 1d shows the predicted SBRE as a function of $\Delta$ for several ER values, computed numerically from Eqs. 2 and \ref{eq9}. For each curve, the imaginary part of $\Delta$ is held at zero, $t$ is fixed at $0.01 - 0.01i$, and $\epsilon_{S} = \epsilon_{BG}/2$. As is anticipated from the diminished degree of coherence, the peak SBRE becomes reduced as ER increases. The degree of reduction is surprisingly severe -- whereas in the ideal case (ER = $0$) an interferometer transmittance of $|\Delta|^{2} = .01$ is expected to yield an enhancement of $10$ for real $\Delta$, the peak enhancement reduces to only 2-3 in a more realistic case where ER = $0.01$.

We are primarily interested in the maximum achievable SBRE and the corresponding optimal $|\Delta|^{2}$, as a function of ER. These can be obtained numerically from Eq. \ref{eq9}, as shown in Fig. 1e (dotted lines). Again considering real $\Delta$ and dropping small terms, we can adopt the compact approximation

\begin{equation} \label{eq11}
    SBRE \approx \frac{\Delta + \epsilon_{S}}{\Delta^{2} + \epsilon_{BG}}
\end{equation}

Setting the first derivative to 0 and continuing to drop small terms, we can estimate the peak $SBRE$ and the optimal value for $\Delta$

\begin{equation} \label{eq12}
    SBRE_{max} \approx \frac{1}{4\sqrt{\text{ER}}}
\end{equation}
\begin{equation} \label{eq13}
    |\Delta|^{2}_{opt} \approx 4 \, \text{ER}
\end{equation}

Comparing the above approximations (Fig. 1e, solid lines) with the more exact results suggests that Eqs. \ref{eq12} and \ref{eq13} provide reasonable rules of thumb for how the efficacy of CBS degrades with ER.
 
Note that Eq. \ref{eq12} is in agreement with the peak enhancement of 2-3 observed in Fig. 1d for ER = $.01$. The factor of $1/4$ in Eq. \ref{eq12} makes clear why the dramatic enhancement predicted by Eq. \ref{eq5} is, in general, much less extreme in practice; for instance, to achieve a SBRE of $10$, Eq. \ref{eq12} imposes an extinction ratio better than 6.25e-4.

\subsection{Imaging Through Birefringent Crystals}

We consider now the implementation of CBS by temporal shearing. In our case, this is achieved with birefringent crystals. A field incident on a crystal is prepared in a sum of the crystal's ordinary and extraordinary polarization eigenstates. The group velocity mismatch for these polarizations results in two cross-polarized fields emerging from the crystal separated in time, with the more delayed being the probe field. After transmission through the sample, the fields must be recombined in time. Conventionally, this is achieved using an additional crystal of the same material and of equal thickness with crossed optic axis. The fields are then added coherently with a linear polarizer, projecting both beams onto the same polarization state. Temporal separations on the order of $10$ ps have been demonstrated, \cite{Dijk2005,Orrit2007,Coleal2024}, which is sufficient for measuring transient absorption lifetimes in many samples \cite{Chen2018,dong2019label,Fischer2016,Huang2018}. 
While the above recombination approach is suitable for a scanning geometry with a single element detector, it presents limitations when applied in a widefield imaging geometry with a camera. As noted in \cite{Perri2019,Candeo2019}, placing the second birefringent crystal near an image plane leads to reduced interferometric contrast where off-axis propagationg angles are present (i.e. sample regions where scattering occurs). Further, placing the second crystal near a Fourier plane produced a spatially varying intensity pattern. Both phenomena are attributable to the angularly-dependent phase shift between the cross-polarized fields in the second crystal; a small ER, then, cannot be maintained throughout the imaging FOV, and the possibility of effective CBS in a widefield geometry becomes compromised. 

To address this problem, we must have an understanding of the phase shift resulting from different propagation angles in a birefringent crystal. Theoretical description has previously been provided using a combination of wave and ray optics, particularly for the case where the crystal's optic axis is orthogonal to the crystal's surface \cite{born2013principles}. A description for the case where the optic axis is parallel to the crystal's surface is given in Appendix A based on diffraction theory \cite{Goodman1996,Mertz2019}, utilizing fewer approximations.

For the case where the fields initially propagate in a medium with refractive index $n_{i} = 1$ and assuming the propagation tilt angles to be small, the difference between accumulated phases for extraordinary and ordinary fields is given by
\begin{equation} \label{phieo}
\begin{split}
    \Delta\phi_{eo}(\hat{\mathbf{k}}_{i};d) &\approx
    \frac{\pi}{\lambda}d  \left[
        (\frac{1}{n_{e}}-\frac{1}{n_{o}})\hat{\mathbf{k}}_{i,x}^{2} +
        (\frac{n_{e}}{n_{o}^{2}}-\frac{1}{n_{o}})\hat{\mathbf{k}}_{i,y}^{2} \right] \\
      & \approx
    \frac{\pi}{\lambda}d  \frac{\Delta n}{n_{o}^{2}}\left[
        -(1-\frac{\Delta n}{n_{o}})\hat{\mathbf{k}}_{i,x}^{2} +
        \hat{\mathbf{k}}_{i,y}^{2}
    \right]
    \end{split}
\end{equation}
where $\Delta n = n_e-n_o$ and $\hat{\mathbf{k}}_{i}$ is a unit vector parallel to $\mathbf{k}_{i}$. Note that an overall constant that does not affect the shape of this function has been dropped, and that this expression is wavelength dependent. For CBS, it will be important to flatten the differential phase dispersion in a broadband manner. As is apparent from Eq. \ref{phieo}, the contours of constant $\Delta\phi_{eo}$ are hyperbolic in shape, with a slight asymmetry between the $x$ and $y$ directions that becomes more prominent for larger $\Delta n/n_{o}$. As an example, for calcite $\Delta n/n_{o}$ is about $10$\%, relatively large for birefringent crystals.

The above results can be validated by experiment. We use ultrafast pulses of center wavelength $810$ nm, bandwidth $30$ nm, and transform limited pulse width of $35$ fs, obtained from an amplified Ti:sapphire laser (Coherent Legend). As described below, dispersion in our system results in an actual pulse width on the order of 100s of fs.
For all experiments, a half-wave plate and quarter-wave plate in series are placed at the start of the optical path, comprising an arbitrary polarization state generator (PSG) \cite{Hariharan1993}; this allows setting any value of $\Delta$ [Fig. 2b, 2c].
Temporal shearing between horizontal and vertical polarizations is then achieved with a calcite plate (Shalom EO) of thickness 16.5 mm placed before the sample. 
The differential group index for cross-polarized fields in calcite is $\Delta n_{g} = -.181$  at $810$ nm\cite{newlightSellmeier}, corresponding to a predicted temporal shear of $\Delta t=|\frac{d}{c}\Delta n_{g}| \approx 10$ ps.
To perform temporal unshearing , we use  two pairs of calcite wedges  (Shalom EO) placed after the sample (the purpose of using two wedge pairs will become apparent below). Each wedge pair is equivalent to a plate whose thickness can be tuned by translating one of the wedges \cite{Perri2019,Candeo2019}. The first and second wedge pairs are adjusted to produce equivalent plate thicknesses of 22.1 mm and 5.6 mm, respectively, such that the (opposing) temporal shears produced by wedge pairs exactly cancel the temporal shear produced by the input plate. The cross-polarized pulses are then recombined interferometrically using a linear polarizer oriented $135^\circ$ from horizontal. 

In a first experiment, a USAF 1951 calibration target was imaged with the two pairs of calcite wedges placed roughly in a conjugate imaging plane [Fig. 2b].
As is evident from Fig. 2d, bright outlines persist about the scattering features in the target, corresponding to regions of reduced interferometric contrast (i.e. poor ER) because of off-axis propagation directions. In a second experiment, the four calcite wedges were placed instead near a Fourier plane, such that the propagation angles out of the wedges (i.e. $\hat{\mathbf{k}}_{i,x}$ and $\hat{\mathbf{k}}_{i,y}$) were mapped directly to position coordinates at the camera sensor [Fig. 2e]. The large intensity variations resulting from non-zero values of $\Delta\phi_{eo}(\hat{\mathbf{k}}_{i};d)$
at off-axis propagation angles  are clearly apparent. As predicted, these variations are hyperbolic in shape, in agreement with results in \cite{Perri2019,Candeo2019}. By adjusting the phase between ordinary and extraordinary polarizations with the PSG, several intensity measurements can be recorded, allowing reconstruction of the phase dependence on off-axis propagation angle [Fig. 2f]. The result of this procedure is in good agreement with the phase predicted by Eq. \ref{phieo} over the same range of angles [Fig. 2g].

\begin{figure}[h]
    \label{crystal_theory_figure}
    \includegraphics[width=13.2cm]{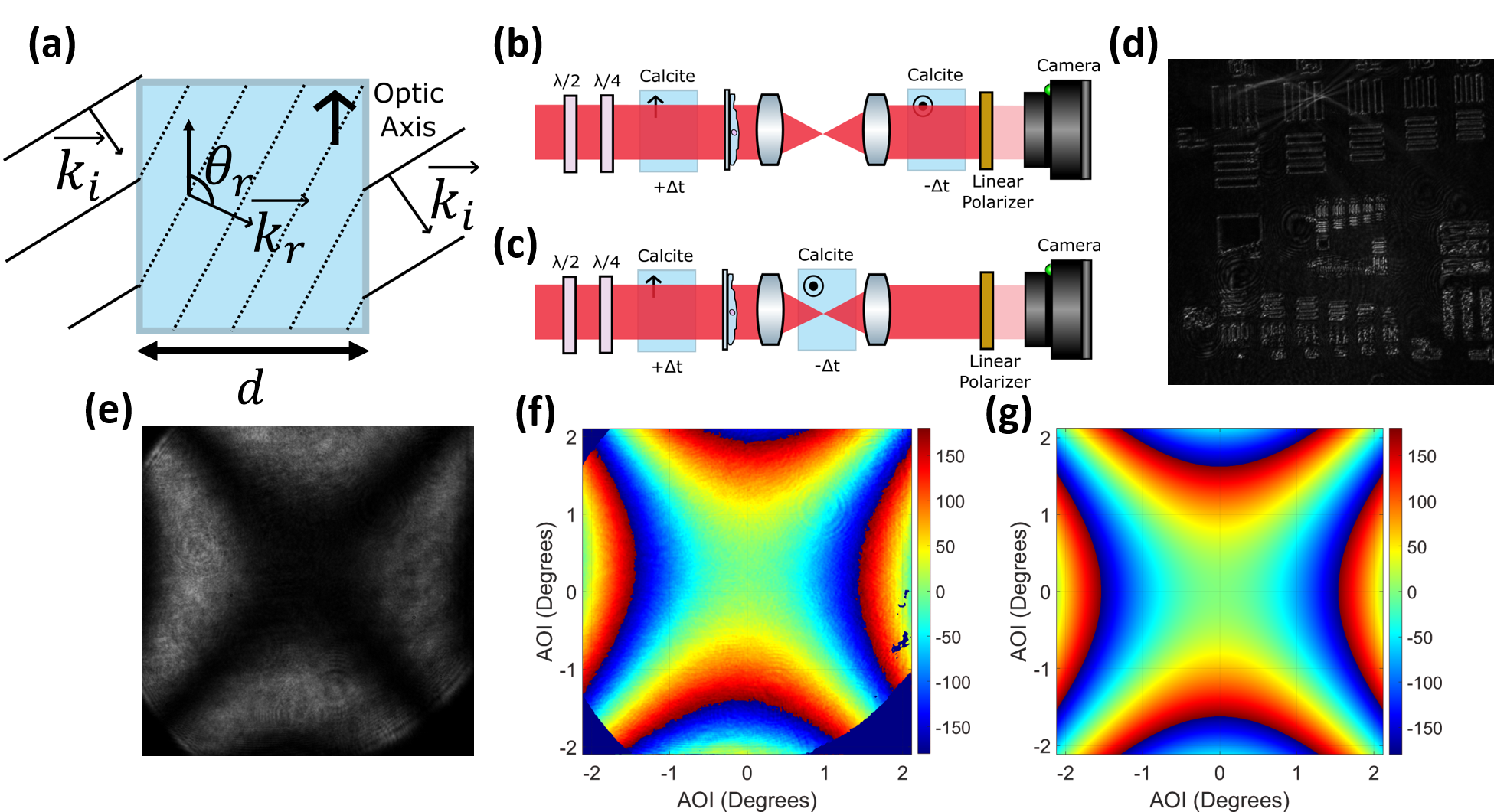}
    \caption{(a) Wavefront diagram depicting off-axis propagation though a uniaxial birefringent crystal. (b) Temporal pulse recombination using a single birefringent crystal 
    near an image plane. (c) As Fig. 2a, but with the crystal near a Fourier plane. 
    (d) Measured image of a USAF resolution target with the crystal in the image plane. Intensity reduction near scattering features is poor.
    (e) Measured intensity distribution with the crystal near the Fourier plane. Intensity reduction varies dramatically over the field of view.
    (f) Measured phase offset with the crystal near the Fourier plane. 
    (g) Predicted phase offset with the crystal near the Fourier plane based on Eq. \ref{phieo}. Note the qualitative agreement with Fig. 2f.}
\end{figure}

\subsection{Angle Compensated Temporal Shearing (ACTS)}

In practice, to implement CBS based on temporal shearing in a widefield imaging configuration, it is necessary to compensate for the dependence of $\Delta\phi_{eo}(\hat{\mathbf{k}}_{i};d)$ on propagation angle. We introduce a method to achieve this. The central idea is to modulate the phase difference between probe and reference beams twice by unshearing in time with two different crystals in series, instead of one. By adjusting different thicknesses for each crystal and placing them in locations along the optical path with different focusing conditions, our interferometer compensates for the angular dependence of $\Delta\phi_{eo}(\hat{\mathbf{k}}_{i};d)$ while continuing to induce an appreciable differential group delay. We call our method Angle-Compensated Temporal Shearing (ACTS). 

A schematic of our method is shown in Fig. 3a.  As before, we use a calcite crystal placed in front of the sample. Since calcite is a negative uniaxial crystal, the field with polarization state orthogonal to the crystal optic axis experiences a larger delay, and thus serves as the probe field. However, the probe field will not have ordinary polarization for every crystal in the system. Generally, whether probe or reference has ordinary polarization will depend on the crystal's orientation and the presence of any waveplates. We therefore consider how the phase difference between probe and reference fields $\Delta\phi_{pr}(\hat{\mathbf{k_{i}}})$ changes at the different positions marked in Fig. 3a. This quantity is distinct from the quantity $\Delta\phi_{pr}(\hat{\mathbf{k}}_{i})$ described in Eq. 12. Since the crossed fields are collimated at input crystal, the phase difference at position 1 is constant over the angular spectrum. This overall constant can be compensated for with the PSG, leading to
\begin{equation}
    \Delta \phi_{pr;1}(\mathbf{\hat{\mathbf{k}}_{i}}) = 0
\end{equation}

The crossed fields then propagate through the sample, and are incident on the first crystal in the ACTS interferometer. If this first crystal's optic axis is oriented parallel with the probe field polarization, the phase difference over the angular spectrum at position 2 is given by Eq.  \ref{phieo}
\begin{equation}
    \Delta \phi_{pr;2}(\hat{\mathbf{k}}_{i}) = \Delta \phi_{eo}(\hat{\mathbf{k}}_{i};d)
\end{equation}

The fields are then imaged to a plane near a second crystal using a 4f relay, where a half-wave plate with fast axis $45^\circ$ from horizontal is placed in the relay. The purpose of this half-wave plate is to swap the polarizations of the probe and reference fields. Crucially, the magnification of this 4f relay is not unity. In our experiment, we use a relay with magnification $M = 1/2$. After magnification, the phase difference over the angular spectrum is given by
\begin{equation}
    \Delta \phi_{pr;3}(\hat{\mathbf{k}}_{i}) = \Delta \phi_{eo}(M\hat{\mathbf{k}}_{i};d)
\end{equation}
where we have used the well known principle that magnification in real space corresponds to de-magnification in Fourier space.

The second crystal in the ACTS interferometer has optic axis parallel to the first crystal's optic axis, and is placed after the 4f relay. Because the polarizations of probe and reference fields are switched in the 4f relay, the phase difference over the angular spectrum induced by this second crystal has opposite sign compared to the first. If the second crystal thickness is set to $M^{2}d$ , the overall phase difference at the output of the ACTS interferometer becomes 
\begin{equation} \label{eq25}
    \Delta \phi_{pr;4}(\hat{\mathbf{k}}_{i}) = \Delta \phi_{eo}(M\hat{\mathbf{k}}_{i};d)
    - \Delta \phi_{eo}(\hat{\mathbf{k}}_{i};M^{2}d)
\end{equation}

We note that the phase differences induced by both crystals largely compensate for one another since $\Delta \phi_{eo}(M\hat{\mathbf{k}}_{i};d)$ is linear in $d$ and, to a good approximation, quadratic in $\hat{\mathbf{k}}_{i}$. That is, the overall phase difference produced by the ACTS interferometer is, to second order, zero. In contrast, the overall differential group delay is 
\begin{equation}
    \Delta \tau = (M^{2} - 1)d \frac{\Delta n}{c}
\end{equation}
Our ACTS interferometer thus provides a nonzero group delay while maintaining a nearly constant phase difference for all angles of incidence, as desired. Crucially, these argument holds regardless of the wavelength dependence noted in Eq. 12; ACTS, then, is spectrally broadband. The choice to employ birefringent wedges in the ACTS interferometer can now be understood, since the wedges allow the crystal thicknesses to be adjusted such that their ratio is $M^{2}$. 

\begin{figure}[h]
    \label{flat_in_k_figure}
    \includegraphics[width=13.2cm]{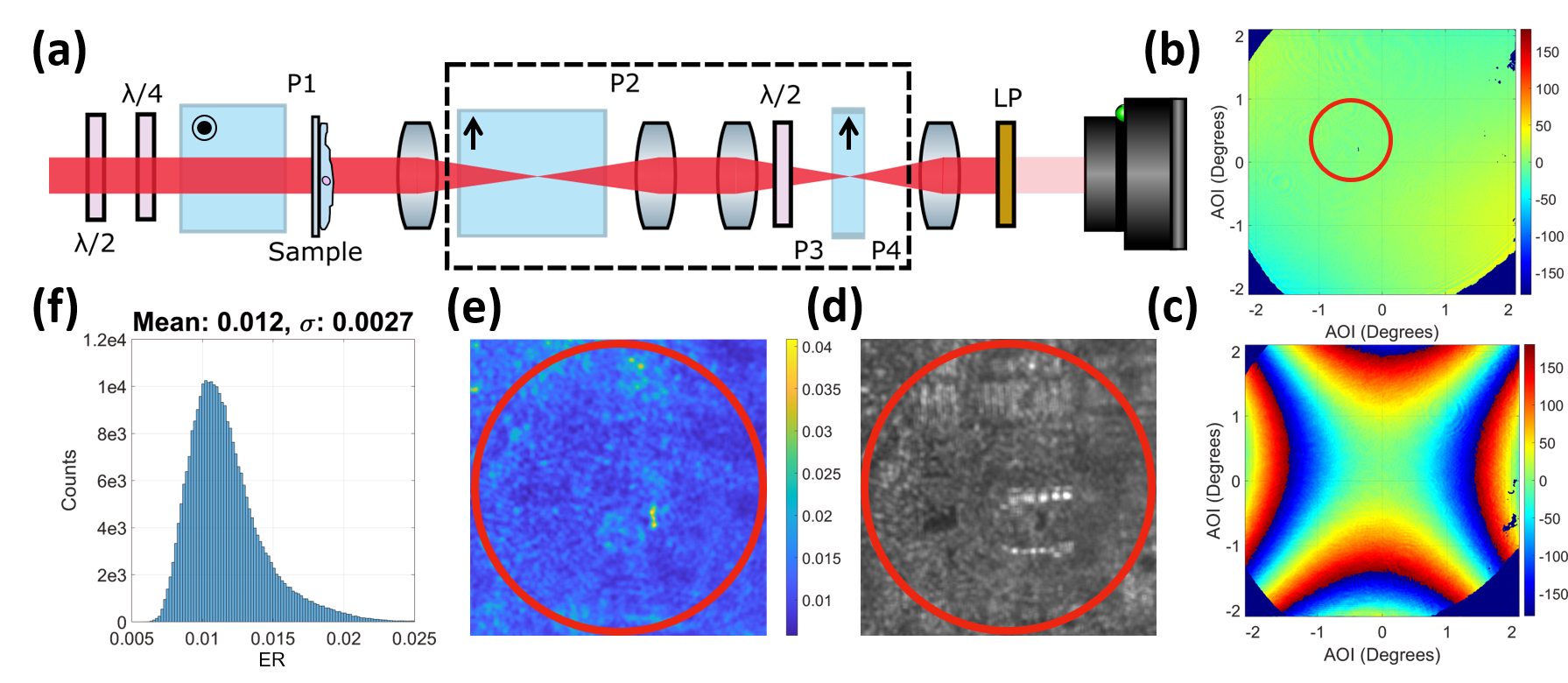}
    \caption{Subfigures ordered clockwise starting from top-left. (a) Simplified sketch of the setup for testing the ACTS interferometer. The dashed box indicated the interferometer itself; points P1 through P4 indicate the locations corresponding to Eqs. 13-16 in the main text. (b) Measured phase offset with the ACTS
    interferometer; the standard deviation of the phase offset is less than $3^\circ$ within the red circle.
    (c) Fig. 2f, the measured phase offset without ACTS, reproduced here for side-by-side comparison with Fig. 3b. (d) Zoomed image of a section of a USAF resolution 
    target within the red circle marked in Fig. 3b under destructive interference. Intensity is amplified $10\times$ to highlight scattering features. (e) Measured extinction ratio within the red circle marked in Fig. 3b. (f) Histogram of ER in the red circle marked in Fig 3b.}
\end{figure}

The proposed ACTS interferometer can be evaluated experimentally. First, we place the ACTS interferometer in a Fourier plane and measure the resultant phase distribution over the angular spectrum, as before [Fig. 3b]. For comparison, the phase distribution without ACTS shown in Fig. 2f is reproduced in Fig. 3c. It is apparent that the phase distribution obtained with ACTS is much flatter than without ACTS. Some residual angular dependence is noted; we attribute this to the challenge of precisely aligning the pitch, roll, and yaw of all four wedges in our ACTS interferometer. Nevertheless, within the red circle marked in Fig. 3b, the phase difference is particularly flat. Fig. 3d shows the intensity distribution near scattering features of a USAF resolution target within this red circle as imaged through the interferometer, where we have enhanced the intensity by a factor of $10$ to facilitate comparison with Fig. 2d. We also estimate the ER in this region, allowing us to predict SBRE. This is done by adjusting the PSG half-wave plate to obtain minimum and maximum intensities, allowing a calculation of the ER at every pixel [Fig. 3e]. The distribution of ER values in the red circle is shown in Fig. 3f. The mean ER is found to be $0.012$, suggesting a predicted SBRE between 2 and 3 based on the curve in Fig. 1e. Though this ER may seem modest compared to extinction ratios commonly achieved with calcite optics, we note that here the calcite crystals are not being used to refract eigenstates to different paths spatially, but rather to split and recombine the eigenstates temporally along the same spatial path. We are thus sensitive to small variations in phase and amplitude accumulated over the optical path. A possible origin of our relatively modest ER and the speckle patterns clearly visible in Fig. 3e is inhomogeneities in the calcite crystals.

Finally, we close this section by recalling that the purpose of our ACTS interferometer is to enable widefield imaging with CBS. We must therefore  relate the parameter $\Delta$ to the experimentally controllable parameters of our interferometer. This problem is treated in Appendix B, leading to the expression 

\begin{equation} \label{delta_of_expt}
    \Delta = 1 + \mathrm{tan}(\theta_{LP})e^{i\phi_{B}}
\end{equation}
where $\theta_{LP}$ is the linear polarizer's roll angle as measured from the horizontal, and $\phi_{B}$ is a phase that can be controlled arbitrarily with the PSG.
Substituting into Eq. \ref{eq4}, we arrive at finally
\begin{equation}
    SBR_{CBS} \approx 2\frac{\left[ 1 + \mathrm{tan}(\theta_{LP})\mathrm{cos}(\phi_{B})\right]\mathrm{Re}\{t\} + 
    \left[\mathrm{tan}(\theta_{LP})\mathrm{sin}(\phi_{B})\right]\mathrm{Im}\{t\}}
    {1 + 2 \mathrm{tan}(\theta_{LP})\mathrm{cos}(\phi_{B}) + \mathrm{tan}^{2}(\theta_{LP})}
\end{equation}
allowing CBS performance to be considered in terms of experimentally accessible parameters as desired. 

\subsection{Widefield Pump-Probe Microscope}

An experimental setup for the implementation of CBS with an ACTS interferometer for a home-built widefield pump-probe microscope is shown in Fig. 4. Pump, probe, and reference beams are derived from the amplified Ti:sapphire laser described in Section 2.3 (repetition rate 1 kHz).

\begin{figure}[h]
    \includegraphics[width=13.2cm]{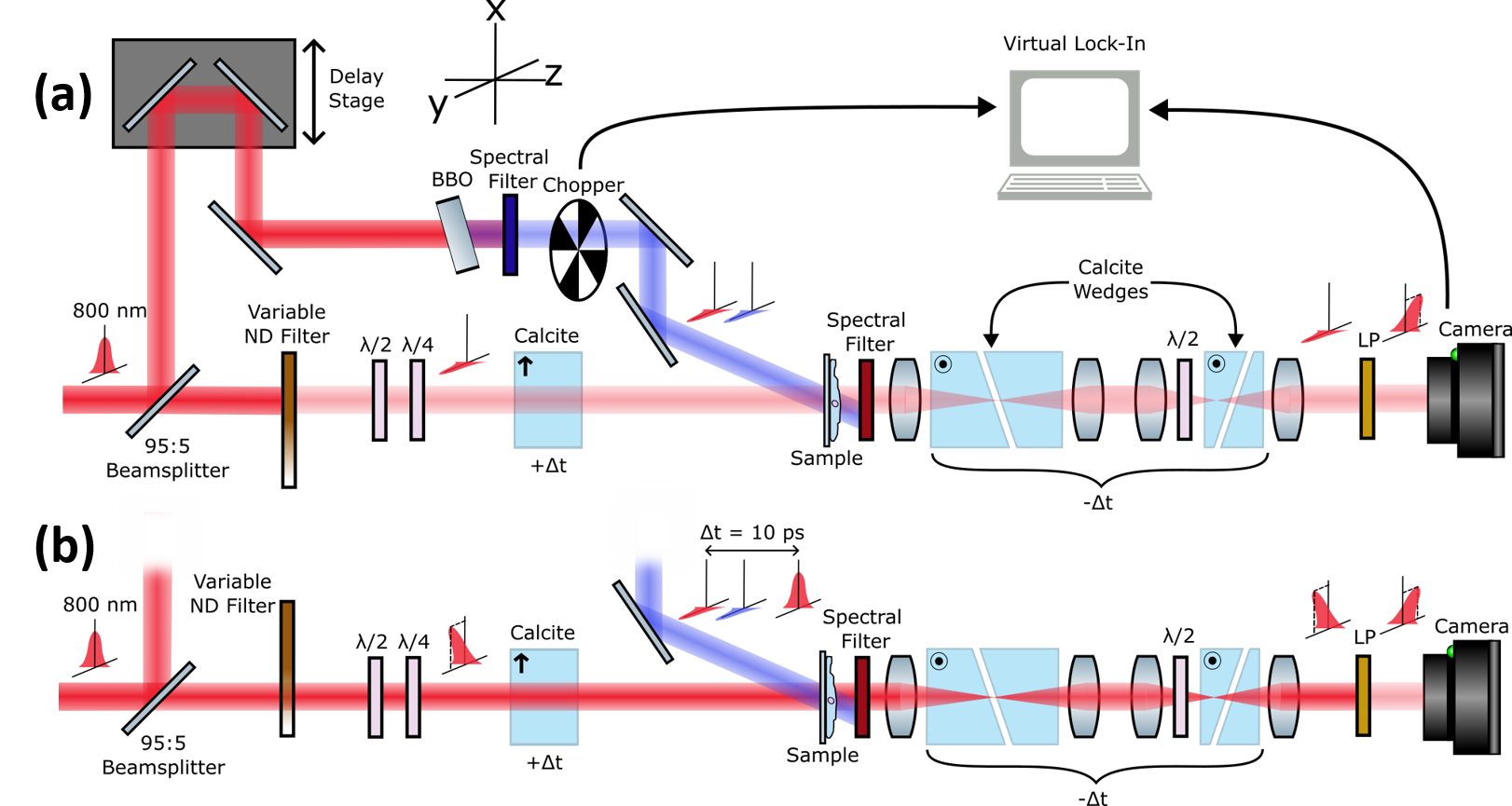}
    \caption{Setup for widefield transient absorption microscopy.  (a) For conventional pump-probe imaging, the PSG is adjusted such that there is no reference field. The variable ND filter is adjusted correspondingly to increase attenuation. The y-axis is orthogonal to the page. (b) When CBS is used, only the imaging arm changes. The PSG is adjusted to produce both probe and reference fields and the variable ND filter's attenuation is decreased.}
\end{figure}

The pump beam is split from the laser's output by a 95:5 beamsplitter before propagating through a variable attenuator comprised of a half-wave plate and two Brewster windows. After attenuation, the pump pulse is frequency doubled by a BBO crystal to generate $405$ nm excitation light. An iris is placed near the BBO crystal to provide control of the pump beam size at the sample. Because of the long propagation distance from the iris to the sample, two imaging relays are used to mitigate diffraction (not shown). To avoid the possibility of self-focusing from the high peak intensities at the focal points in these relays, a cylindrical lens of focal length -1000 mm is inserted immediately before the BBO crystal (also not shown). The long focal length ensures the beam is only weakly perturbed at the sample, but is sufficient to reduce peak intensities by stretching the beam profile into a line. To maximize signal, the pump beam's polarization is set to be co-polarized with the probe beam's polarization.

Probe and reference beams are derived from the 5\% of light reflecting from the 95:5 beamsplitter, and are temporally sheared as described above. After transmission through the sample, they are collected by a 50 mm focal length lens and recombined by the ACTS interferometer. The first lens in the interferometer has a focal length of $200$ mm and the second $100$ mm. The effective thickness of one wedge pair is adjusted until the extinction ratio is as small as possible, indicating that the optical path difference  between probe and reference beams is much smaller than the beam coherence length. To further achieve the flattest interference pattern possible, the effective thicknesses of the wedge are tuned simultaneously, while adjusting the pitch, roll, and yaw of all four wedges.

Scanning of the pump pulse arrival time relative to the probe and reference pulses is achieved using a motorized delay stage in the pump path. To detect the pump-probe signal, we adopt the strategy of virtual lock-in detection \cite{Bai2019}, where we rapidly modulate the pump beam and subtract intensity images with the pump transmitting and blocked.
A chopper placed in the pump beam path performs the rapid modulation. The 31st subharmonic of the Q-switch clock from the Ti:sapphire source's seed laser is used to synchronize the camera, chopper, and laser at 32.3 Hz. To ensure the camera can operate at this frequency, we restrict the camera's exposure time to 1 ms, corresponding to capturing a single pulse per sensor readout. We use a global shutter camera to ensure that the spatiotemporal intensity variations due to RIN originate from the same pulses at every pixel for each readout. To compensate for these variations, we make use of a self-referencing approach described in Ref. \cite{Hormann2024}. This approach entails estimating the RIN-induced variations within the field of view (the image region spanned by the pump beam) based on an interpolation from measurements acquired outside the field of view. In Ref. \cite{Hormann2024}. this interpolation was linear. Here, we use instead a quadratic interpolation, leading to an overall 5-6 times reduction in RIN-induced noise. 

To further increase SNR, we use spatiotemporal averaging. In the transient absorption images shown below, spatial averaging is performed using sliding windows $10 \times 10$ pixels in size (corresponding to 18 $\mu$m $\times$ 18 $\mu$m at the sample -- for reference, our diffraction limited resolution is about 2 $\mu$m). Temporal averaging varies between 1 to $240$ frames for different experiments, as indicated below.

The temporal resolution of our transient absorption measurements is defined, in principle, by the pulse-width of our laser beams, which has a bandwidth of 30 nm and corresponding nominal transform limited pulse duration of 35 fs. However, because we made no special effort to compensate for dispersion, the true pulse width is expected to be longer. To evaluate our system's temporal resolution, we adjusted the probe and reference beams to have equal amplitudes with a $\pi/2$ phase shift, allowing phase sensitive measurements. The Kerr effect (assumed instantaneous) was then measured in a glass slide, providing a measure of the convolution of the probe and pump beams. This procedure yielded an estimate for the temporal resolution of $260$ fs, which we take as our temporal resolution.

\section{Results and Discussion}

Our goal is to compare CBS to non-interferometric microscopy, which is only sensitive to the real part of $t$.
For this, we intentionally adjust our PSG so that our system has no sensitivity to the imaginary part of $t$ ( i.e., we set the imaginary part of $\Delta$ to zero). This is equivalent to setting $\phi_{B} = 0$, which we achieve by rotating the analyzer $135^\circ$ and adjusting the PSG's half-wave plate to an intensity minimum. To ensure $\Delta$ has a finite real part, the analyzer is displaced slightly from this minimum position. Based on our measured ER of $.012$ [Fig. 3f] and making use of Eq. \ref{eq12}, we expect that an analyzer rotation of $142^\circ$ should maximize SBRE. However, larger rotation angles have the benefit of reducing the effects of speckle noted in Fig. 3e. For this reason, we generally set the analyzer rotation to $147^\circ$, which represents a good compromise between enhancement and intensity homogeneity. This operating point is used for our demonstrations of CBS below, unless otherwise noted. 

To compare with results obtained in the absence of CBS, we simply remove the reference field. This is done in several steps. First, the analyzer transmission axis is set to vertical; because of the half-wave plate in the ACTS interferometer, this results in an intensity minimum when the field emerging from the PSG has purely vertical (probe) polarization.
Next, the PSG quarter-wave plate is rotated so its fast axis is vertical,
resulting in an elliptical state with vertical azimuth. 
Finally, the PSG half-wave plate is rotated, modulating the state's ellipticity until an intensity minimum is achieved and producing a vertically polarized state as desired. Finally, the analyzer is rotated back to its initial position used when recording data with CBS to minimize experimental differences between the two measurements. During this process, the neutral density filter is adjusted such that the intensity at the camera is the same with and without CBS, for comparison of the two modes under equal detection conditions (i.e same shot noise level, readout noise, etc.).

For our demonstrations, we use gold nanorods (AuNRs), which have a large, well characterized transient absorption response near the longitudinal resonance \cite{Dijk2005}.  AuNR films were prepared by drop-casting AuNR solutions (NR dimensions $100$ nm in length and $30$ nm in width) on glass coverslips. Darkfield microscopy of these films (not shown) reveals boundaries between regions, which will serve as structural features in our transient absorption images. 

To begin, we measure the decay of the transient absorption signal with and without CBS. Selected images are shown in Fig. 5a. The circular region where signal is apparent corresponds to the shape of the pump beam. Because the pump beam illuminates the sample from an oblique angle, this region shifts over time, appearing to "roll" across the field of view. Note that the same region boundaries revealed with darkfield microscopy images are also apparent here. To better visualize the transient absorption decay, we plot time traces at locations in and out of the pump region [Fig. 5b]. For this experiment, and for the others described below, 240 pairs of hot and cold frames were averaged for each image. As is evident from these traces, the SBR is clearly enhanced with CBS. 

\begin{figure}[h]
    \label{time_series_figure}
    \includegraphics[width=13.2cm]{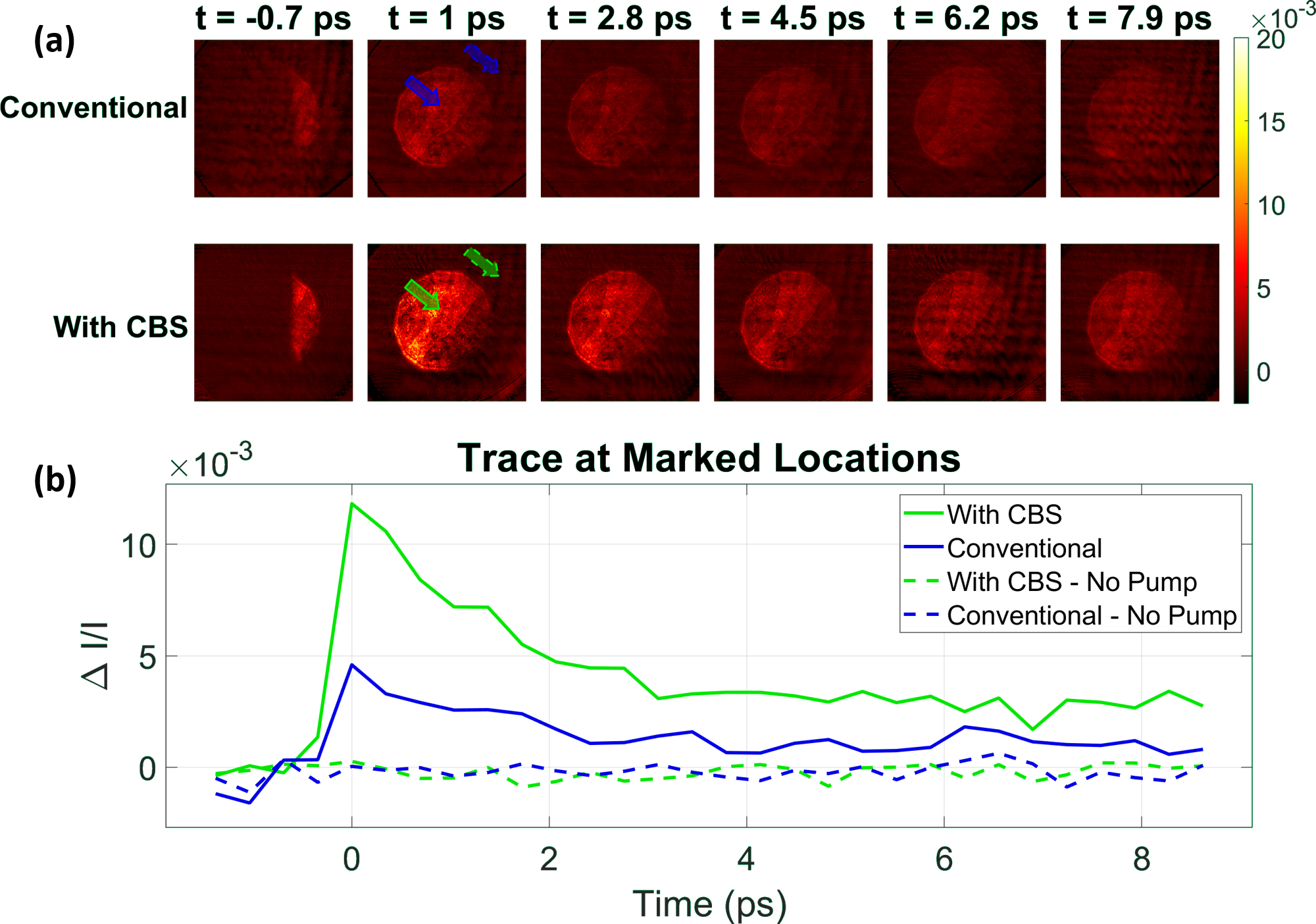}
    \caption{(a) Selected transient absorption images recorded without (conventional) and with CBS. (b) Time traces obtained from the locations indicated by arrows; the time axis is shifted such that peaks occur at t=0.}
\end{figure}

To quantify SBRE and SNRE, we record 15 measurements with the delay stage kept stationary at a position where the pump beam has just finished rolling across the FOV. The signal mean and standard deviation are then computed at every pixel in the field of view. SBRE is computed as the pixel-wise ratio of virtual lock-in signals with and without CBS [Fig. 6a]. A histogram of SBRE values inside the red circle corresponding to well-compensated phase [see Fig. 3b] shows clear SBR enhancement with CBS, in agreement with expected values [Fig. 6b]. Further, a 2D histogram of the mean signals with and without CBS reveals a strong linear correlation, indicating that the interferometer has enhanced SBR without distorting the image within the red circle [Fig. 6c]. Similar calculations are repeated for SNR, computed as the pixel-wise ratio of mean and standard deviation for the virtual lock-in signal [Fig. 6d-f]. Note that the mean SNRE is lower than the mean SBRE; we attribute this to additional sensitivity to noise in the pump beam introduced by CBS.
Nevertheless, the mean SNRE is 1.7. In principle, since SNR scales as the square root of temporal averaging time, this allows a nearly $3\times$ enhancement of temporal resolution.

\begin{figure}[h]
    \label{Enhancement_demo_figure}
    \includegraphics[width=13.2cm]{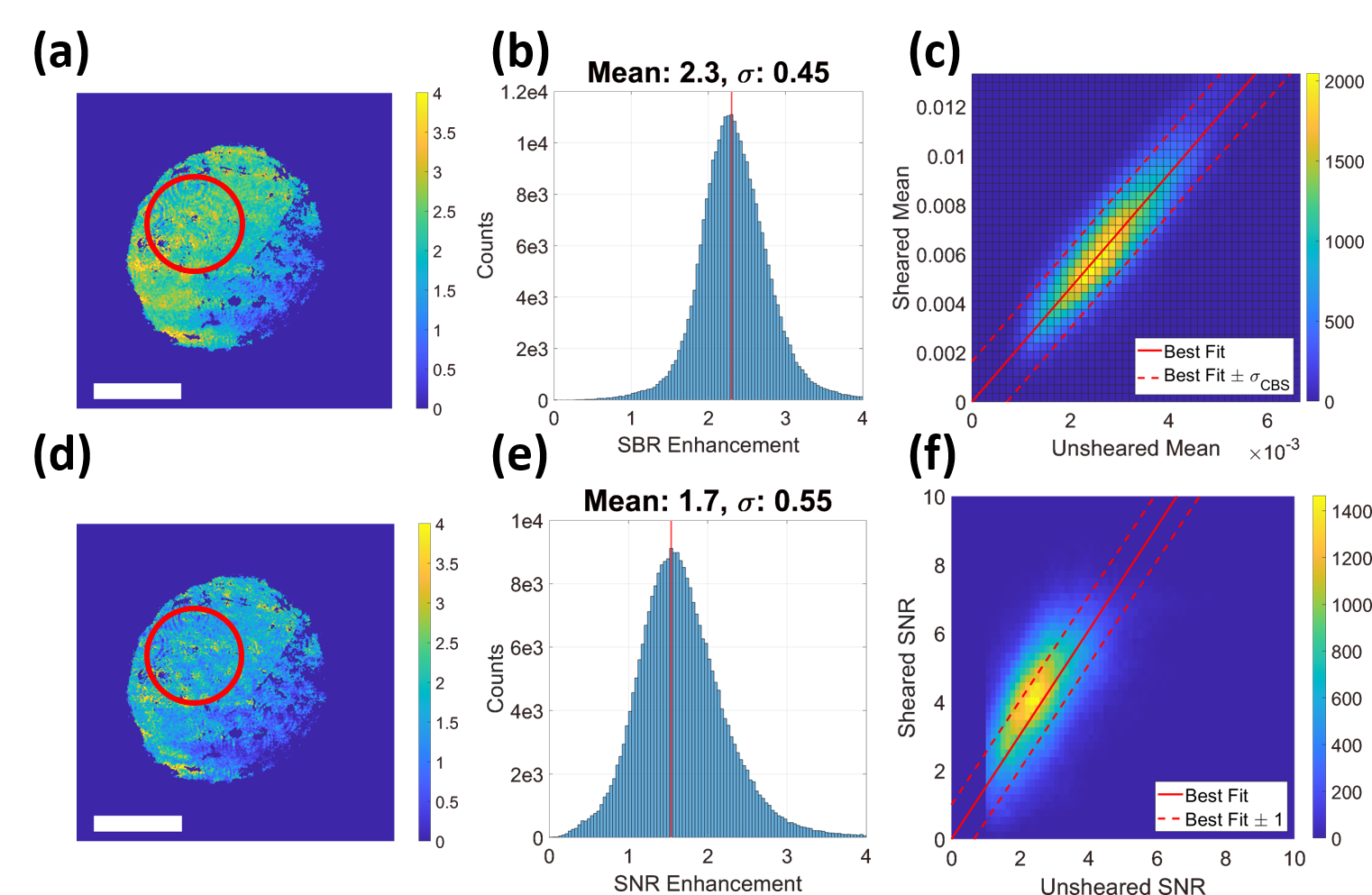}
    \caption{(a) Measured SBRE. The indicated red circle corresponds with a region of relatively small phase offset variation, i.e., successful ACTS. (b) Histogram of SBRE values inside 
    the red circle in Fig. 6a. (c) 2D histogram of mean transient-absorption signals; $\sigma_{CBS}$ is taken from the standard deviation noted in Fig. 6b. (d) Measured SNRE. (e) Histogram of SNRE values inside the red circle in Fig. 6d. (f) 2D histogram of SNR values.}
\end{figure}

Finally, we experimentally interrogate the dependence of SBRE on real $\Delta$ as predicted by Eq. \ref{eq11}. We do so by varying the linear polarizer rotation angle by $2^\circ$ increments about $135^\circ$; we call this bias angle away from $\theta_{B} = \theta_{LP} - 135^{\circ}$. As the analyzer rotation is varied, the variable neutral density filter is adjusted to maintain a constant intensity at the camera. Fig. 7a shows SBRE images within the red circle for selected analyzer angles. As expected, the sign of the SBRE flips near $\theta_{B} = 0$. Note that patterns of positive and negative SBRE are evident in the $\theta_{B} = 0$ image displaying correlation with diffraction rings, suggesting that scattering features in the optical path may cause variations in polarization state rotation or relative phase shift that can perturb the interferometer's operating point. Histograms of SBRE values inside the red circle for the images selected in Fig. 7a are shown in Fig. 7b. The mean and standard deviation of these histograms are plotted in Fig. 7c, along with a fit using Eqs. \ref{eq11} and 18. This fit yielded estimated values for $\epsilon_{s}$ and $\epsilon_{BG}$ of $.06$ and $.05$, respectively. From Eq. \ref{eq10}, this corresponds to a fitted ER of $.01$, consistent with the measured value. These experiments were performed by averaging $240$ pairs of hot and cold frames per analyzer angle.

\begin{figure}[h]
    \label{enhancement_curve_figure}
    \includegraphics[width=13.2cm]{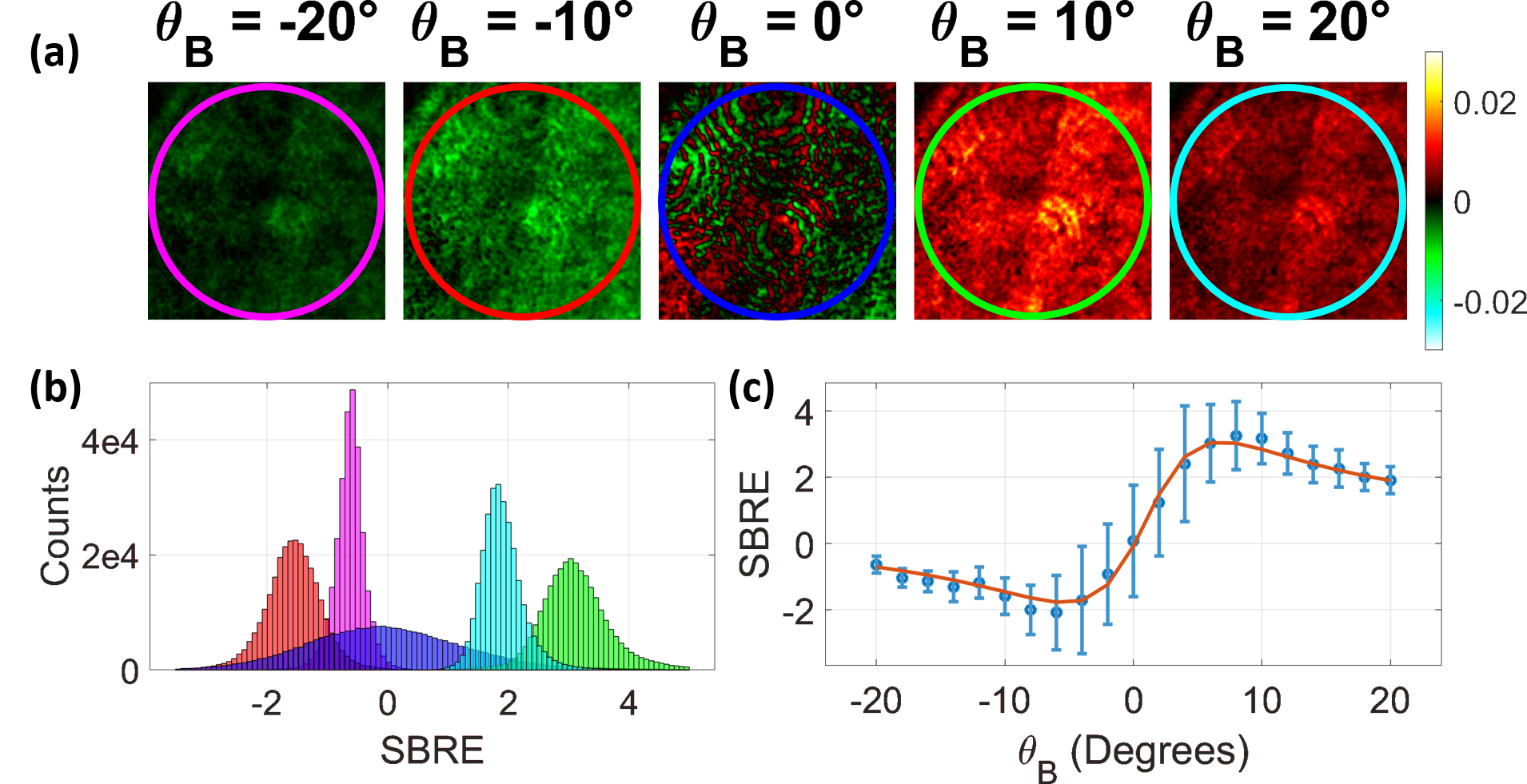}
    \caption{(a) SBRE images for various analyzer rotation angles. (b) Histograms of
    SBRE values inside the corresponding colored circles in Fig. 7a. (c) SBRE enhancement as a function of analyzer rotation angle. Means and standard deviations are derived from histograms like those in Fig. 7b. Solid line is theoretical fit based on Eq. \ref{eq11}.}
\end{figure}

\section{Conclusion}

Two novel ideas have been demonstrated. First, we described a method of CBS to enhance imaging contrast in systems limited by camera saturation. Second, to enable CBS in a widefield imaging geometry, we introduced a novel ACTS interferometer. To our knowledge, this represents the first demonstration of an interferometer with compensated dependence on $\mathbf{k}_{i}$. By applying CBS and ACTS in combination, enhanced SNR in pump-probe imaging was demonstrated.

At present, our SBR enhancement remains limited by the ER achievable with the ACTS interferometer. We speculate that the relatively modest ER attainable is due to inhomogeneities in the calcite crystals. Greater ER could be potentially achieved with a different material, such as BBO. The degree of compensation for dependence on angle of incidence is also currently somewhat less than anticipated, likely due to the challenge of precisely aligning all four wedges in the ACTS interferometer. 
This could be improved by constructing the interferometer more simply from plates instead.

We observed an SNRE somewhat smaller than the SBRE, which we attribute to noise correlated with the pump beam. For this reason, the contrast enhancement is currently limited to regimes where noise other than this pump noise is dominant, e.g. shot noise or RIN in the probe and reference beams. In principle, this noise could be compensated by monitoring the pump-beam intensity separately with a photodiode, thus extending the contrast enhancement of CBS to a wider 
range of experimental conditions.

Finally, we note that, because our temporal shearing approach makes use of cross-polarized beams, a small ER is only achievable for samples that exhibit weak (ideally negligible) birefringence. For such samples, CBS combined with an optimized ACTS interferometer can dramatically increase imaging speed by reducing the need for frame averaging. As an example, this could be beneficial for industrial quality inspection of graphene films, whose anticipated birefringence is small. More generally, as a module uniquely insensitive to propagation direction, the ACTS interferometer can be attractive for a variety of applications beyond widefield pump-probe microscopy, such as hyperspectral imaging or phase-sensitive laser scanning microscopy.

\section*{Appendix A: Angle-Dependent Differential Phase}

Here, we provide a theoretical description of the differential phase accumulated between fields with ordinary and extraordinary polarization in terms of incident $\hat{\mathbf{k}}$ vectors, i.e., Fourier optics.

To begin, we consider a plane wave with wavevector $\mathbf{k}_{i}$ and direction $\hat{\mathbf{k}}_{i}=\mathbf{k}_{i}/|\mathbf{k}_{i}|$ initially traveling in an isotropic medium of refractive index $n_{i}$ [Fig. 2a]. The wave is incident on a uniaxial birefringent medium with thickness $d$, parallel faces, and optic axis oriented along the $y$ axis. In the birefringent medium, the wave propagates in a new direction $\hat{\mathbf{k}}_{r}$ with associated refractive index $n_r$, both of which depend on $\hat{\mathbf{k}}_{i}$, the incident field's polarization, and the birefringent medium's ordinary and extraordinary refractive indices, $n_{o}$ and $n_{e}$. The necessary conditions for phase matching at the interface are
\begin{equation} \label{eq17} \tag{A1}
    \mathbf{k}_{i,||} = \mathbf{k}_{r,||} \rightarrow n_{i}\hat{\mathbf{k}}_{i,||} = n_{r}
    \hat{\mathbf{k}}_{r,||}
\end{equation}
where the subscript $||$ indicates the component of the vector parallel to the interface. Since the birefringent medium has parallel faces, the wave exits with wavevector parallel to the incident wavevector, regardless of the incident field's polarization state relative to the optic axis. In other words, the field's angular spectrum experiences no deformation of its modulus. It remains to determine the accumulated phase over the angular spectrum for ordinary and extraordinary polarization states when propagating over the distance $d$.

For both states, this accumulated phase can be written as (making use of Eq. \ref{eq17})
\begin{equation} \label{eq18} \tag{A2}
    \phi =  d \, \mathbf{k}_{r} \cdot \hat{\mathbf{z}}=
    d\frac{2\pi}{\lambda}n_{r}\hat{\mathbf{k}}_{z}=
    \frac{2 \pi}{\lambda}dn_{r}
    \sqrt{1 - \frac{n_{i}^{2}}{n_{r}^{2}} \hat{\mathbf{k}}_{i,x}^{2}
    - \frac{n_{i}^{2}}{n_{r}^{2}} \hat{\mathbf{k}}_{i,y}^{2}}
\end{equation}

The problem now reduces to finding $n_{r}$ for different polarization states. For fields with ordinary polarization, $n_{r} = n_{o}$ for all $\hat{\mathbf{k}}_{i}$, and the accumulated phase follows immediately by substitution

\begin{equation} \tag{A3}
    \phi_{o} = \frac{2 \pi}{\lambda}dn_{o}
    \sqrt{1 - \frac{n_{i}^{2}}{n_{o}^{2}} \hat{\mathbf{k}}_{i,x}^{2}
    - \frac{n_{i}^{2}}{n_{o}^{2}} \hat{\mathbf{k}}_{i,y}^{2}}
\end{equation}

For waves with extraordinary polarization, the solution is more involved. The refractive index $n_{r}$ now depends on the angle $\theta_{r}$ [Fig. 2a] between $\hat{\mathbf{k}}_{r}$ and the optic axis according to  \cite{Reider2016}
\begin{equation} \tag{A4}
    \frac{1}{n_{r}^{2}(\theta_{r})} = 
    \frac{\mathrm{cos}^{2}(\theta_{r})}{n_{o}^{2}} + 
    \frac{\mathrm{sin}^{2}(\theta_{r})}{n_{e}^{2}}
\end{equation}

Recalling that the crystal optic axis is oriented along the $y$ axis, the trigonometric terms above can be expressed in terms of $\hat{\mathbf{k}}_{r}$

\begin{equation} \tag{A5}
    \begin{split}
    \mathrm{cos}^{2}(\theta_{r}) &= 
    (\hat{\mathbf{k}}_{r} \cdot \hat{\mathbf{y}})^{2} = \hat{\mathbf{k}}_{r,y}^{2} \\
    \mathrm{sin}^{2}(\theta_{r}) &= |\hat{\mathbf{k}}_{r} \times \hat{\mathbf{y}}|^{2} = \hat{\mathbf{k}}_{r,x}^{2}
     + \hat{\mathbf{k}}_{r,z}^{2}
    \end{split}
\end{equation}

Eqs. A4 and A5 then yield:

\begin{equation} \tag{A6}
    n_{r}\hat{\mathbf{k}}_{z} = 
    n_{e}\sqrt{1 - \frac{n_{r}^{2}\hat{\mathbf{k}}_{r,y}^{2}}{n_{o}^{2}} - 
    \frac{n_{r}^{2}\hat{\mathbf{k}}_{r,x}^{2}}{n_{e}^{2}}}
\end{equation}

The right-hand side can be related to $\hat{\mathbf{k}}_{i}$ by Eq. A1. We find then, with Eqs. \ref{eq17}, \ref{eq18}, and A6
\begin{equation} \tag{A7}
    \phi_{e} = \frac{2 \pi}{\lambda}dn_{e}\sqrt{1 - 
    \frac{n_{i}^{2}}{n_{e}^{2}}\hat{\mathbf{k}}_{i,x}^{2} -
    \frac{n_{i}^{2}}{n_{o}^{2}}\hat{\mathbf{k}}_{i,y}^{2}
    }
\end{equation}

Taking the difference between Eqs. A3 and A7 leads to Eq. \ref{phieo} in the main text.

\section*{Appendix B: Relating $\Delta$ to Experimental Parameters}

To calculate $\Delta$ from experimentally controllable parameters, we again consider the propagation of the fields through the system, but this time treat the ACTS interferometer as a unit and track the evolution the polarization state encoding probe and reference fields. 

The natural formalism for describing cross-polarized fields is Jones calculus. Immediately after the PSG, the vectorial field is proportional to
\begin{equation} \tag{B1}
    E \propto \begin{pmatrix}
        e^{i\phi_{B}} \\
        1
    \end{pmatrix}
\end{equation}
where, again, the PSG quarter-wave plate has fast axis oriented $45^\circ$ from horizontal, and $\phi_{B}$ is a phase bias that can be adjusted arbitrarily with the PSG. This field becomes sheared in time, propagates through a sample of spatially varying complex transmittance, and is then recombined in time by the ACTS interferometer. Neglecting the effects of de-focusing, the field immediately before the linear polarizer is given by
\begin{equation} \tag{B2}
    E \propto \begin{pmatrix}
        1 + t \\
        e^{i\phi_{B}}
    \end{pmatrix}
\end{equation}
where $t$ is the change in sample complex transmittance induced by the pump, as described above. Note that the polarization of the field carrying the phase bias has changed due to the half-wave plate in the ACTS interferometer. Employing a useful factorization for the Jones matrix of a linear polarizer, the field at the camera can be written as:
\begin{equation} \tag{B3}
    E_{cam} \propto 
    \begin{pmatrix}
        \mathrm{cos}(\theta_{LP}) \\
        \mathrm{sin}(\theta_{LP})
    \end{pmatrix}
    \begin{pmatrix}
        \mathrm{cos}(\theta_{LP}) &
        \mathrm{sin}(\theta_{LP})
    \end{pmatrix}
    \begin{pmatrix}
        1 + t \\
        e^{i\phi_{B}}
    \end{pmatrix}
\end{equation}
where $\theta_{LP}$ is the orientation of the linear polarizer's transmitting axis from horizontal. The first column vector is simply a unit vector describing the field's direction and can be omitted from a description of the field's amplitude. We have then
\begin{equation} \tag{B4}
    E_{cam} \propto \mathrm{cos}(\theta_{LP})(1 + t) + 
    \frac{E_{0}}{\sqrt{2}}\mathrm{sin}(\theta_{LP})e^{i\phi_{B}}
\end{equation}
which is readily interpretable as a combination of a modulated probe field $E_{probe} \propto \frac{E_{0}}{\sqrt{2}}\mathrm{cos}(\theta_{LP})$ and a reference field $E_{ref} \propto -\frac{E_{0}}{\sqrt{2}}\mathrm{sin}(\theta_{LP})e^{i\phi_{B}}$. Comparison with Eq. \ref{eq3} leads to the expression for $\Delta$ in Eq. \ref{delta_of_expt} in the main text.

\begin{backmatter}
\bmsection{Funding}
This work was partially funded by the Boston University Photonics Center and the National Institutes of Health (R01NS116139).

\bmsection{Disclosures}
The authors declare no conflicts of interest.

\bmsection{Data availability} Data underlying the results presented in this paper are not publicly available at this time but may be obtained from the authors upon reasonable request.
\end{backmatter}


\bibliography{CBS_Paper}

\begin{thebibliography}{10}
\newcommand{\enquote}[1]{``#1''}

\bibitem{Fischer2016}
M.~C. Fischer, J.~W. Wilson, F.~E. Robles, and W.~S. Warren, \enquote{Invited
  review article: Pump-probe microscopy,} {\protect\JournalTitle{Review of
  Scientific Instruments}} \textbf{87}, 31101 (2016).

\bibitem{Hu2019}
F.~Hu, L.~Shi, and W.~Min, \enquote{Biological imaging of chemical bonds by
  stimulated raman scattering microscopy,} {\protect\JournalTitle{Nature
  Methods}} \textbf{16}, 830--842 (2019).

\bibitem{Zhu2020}
Y.~Zhu and J.~X. Cheng, \enquote{Transient absorption microscopy: Technological
  innovations and applications in materials science and life science,}
  {\protect\JournalTitle{Journal of Chemical Physics}} \textbf{152}, 20901
  (2020).

\bibitem{dong2017pump}
P.-T. Dong and J.-X. Cheng, \enquote{Pump--probe microscopy: theory,
  instrumentation, and applications,} {\protect\JournalTitle{Spectroscopy}}
  (2017).

\bibitem{Zhang2014}
D.~Zhang, P.~Wang, M.~N. Slipchenko, and J.~X. Cheng, \enquote{Fast vibrational
  imaging of single cells and tissues by stimulated raman scattering
  microscopy,} {\protect\JournalTitle{Accounts of Chemical Research}}
  \textbf{47}, 2282--2290 (2014).

\bibitem{Zhang2016}
D.~Zhang, C.~Li, C.~Zhang, \emph{et~al.}, \enquote{Depth-resolved mid-infrared
  photothermal imaging of living cells and organisms with submicrometer spatial
  resolution,} {\protect\JournalTitle{Science Advances}} \textbf{2} (2016).

\bibitem{Virgili2012}
T.~Virgili, G.~Grancini, E.~Molotokaite, \emph{et~al.}, \enquote{Confocal
  ultrafast pump-probe spectroscopy: A new technique to explore nanoscale
  composites,} {\protect\JournalTitle{Nanoscale}} \textbf{4}, 2219--2226
  (2012).

\bibitem{Li2015}
J.~Li, W.~Zhang, T.~F. Chung, \emph{et~al.}, \enquote{Highly sensitive
  transient absorption imaging of graphene and graphene oxide in living cells
  and circulating blood,} {\protect\JournalTitle{Scientific Reports}}
  \textbf{5} (2015).

\bibitem{Chen2018}
A.~J. Chen, X.~Yuan, J.~Li, \emph{et~al.}, \enquote{Label-free imaging of heme
  dynamics in living organisms by transient absorption microscopy,}
  {\protect\JournalTitle{Analytical Chemistry}} \textbf{90}, 3395--3401 (2018).

\bibitem{Dong2019}
P.~T. Dong, H.~Lin, K.~C. Huang, and J.~X. Cheng, \enquote{Label-free
  quantitation of glycated hemoglobin in single red blood cells by transient
  absorption microscopy and phasor analysis,} {\protect\JournalTitle{Science
  Advances}} \textbf{5} (2019).

\bibitem{Huang2018}
K.~C. Huang, J.~McCall, P.~Wang, \emph{et~al.}, \enquote{High-speed
  spectroscopic transient absorption imaging of defects in graphene,}
  {\protect\JournalTitle{Nano Letters}} \textbf{18}, 1489--1497 (2018).

\bibitem{Tipping2016}
W.~J. Tipping, M.~Lee, A.~Serrels, \emph{et~al.}, \enquote{Stimulated raman
  scattering microscopy: an emerging tool for drug discovery,}
  {\protect\JournalTitle{Chemical Society Reviews}} \textbf{45}, 2075--2089
  (2016).

\bibitem{zong2021background}
H.~Zong, C.~Yurdakul, Y.~Bai, \emph{et~al.}, \enquote{Background-suppressed
  high-throughput mid-infrared photothermal microscopy via pupil engineering,}
  {\protect\JournalTitle{ACS photonics}} \textbf{8}, 3323--3336 (2021).

\bibitem{Matthews2011}
T.~E. Matthews, I.~R. Piletic, M.~A. Selim, \emph{et~al.}, \enquote{Pump-probe
  imaging differentiates melanoma from melanocytic nevi,}
  {\protect\JournalTitle{Science Translational Medicine}} \textbf{3} (2011).

\bibitem{Tyler2015}
D.~S. Tyler, F.~E. Robles, M.~C. Fischer, \emph{et~al.}, \enquote{Pump-probe
  imaging of pigmented cutaneous melanoma primary lesions gives insight into
  metastatic potential,} {\protect\JournalTitle{Biomedical Optics Express, Vol.
  6, Issue 9, pp. 3631-3645}} \textbf{6}, 3631--3645 (2015).

\bibitem{Fantuzzi2023}
E.~M. Fantuzzi, S.~Heuke, S.~Labouesse, \emph{et~al.}, \enquote{Wide-field
  coherent anti-stokes raman scattering microscopy using random illuminations,}
  {\protect\JournalTitle{Nature Photonics}} \textbf{17}, 1097--1104 (2023).

\bibitem{Bai2019}
Y.~Bai, D.~Zhang, L.~Lan, \emph{et~al.}, \enquote{Ultrafast chemical imaging by
  widefield photothermal sensing of infrared absorption,}
  {\protect\JournalTitle{Science Advances}} \textbf{5} (2019).

\bibitem{Matthews2017}
T.~E. Matthews, I.~R. Piletic, M.~A. Selim, \emph{et~al.},
  \enquote{Three-dimensional wide-field pump-probe structured illumination
  microscopy,} {\protect\JournalTitle{Optics Express, Vol. 25, Issue 7, pp.
  7369-7391}} \textbf{25}, 7369--7391 (2017).

\bibitem{Massaro2016}
E.~S. Massaro, A.~H. Hill, and E.~M. Grumstrup, \enquote{Super-resolution
  structured pump-probe microscopy,} {\protect\JournalTitle{ACS Photonics}}
  \textbf{3}, 501--506 (2016).

\bibitem{Hormann2024}
M.~Hörmann, F.~Visentin, S.~K. Chakraborty, \emph{et~al.},
  \enquote{Self-referencing for quasi shot-noise-limited widefield transient
  microscopy,} {\protect\JournalTitle{Optics Express, Vol. 32, Issue 12, pp.
  21230-21242}} \textbf{32}, 21230--21242 (2024).

\bibitem{Yaqoob2016}
Z.~Yaqoob, P.~Hosseini, R.~Zhou, \emph{et~al.}, \enquote{Pushing phase and
  amplitude sensitivity limits in interferometric microscopy,}
  {\protect\JournalTitle{Optics Letters, Vol. 41, Issue 7, pp. 1656-1659}}
  \textbf{41}, 1656--1659 (2016).

\bibitem{Dijk2005}
M.~A.~V. Dijk, M.~Lippitz, and M.~Orrit, \enquote{Detection of acoustic
  oscillations of single gold nanospheres by time-resolved interferometry,}
  {\protect\JournalTitle{Physical Review Letters}} \textbf{95} (2005).

\bibitem{Orrit2007}
M.~Orrit, M.~A. van Dijk, M.~Lippitz, and D.~Stolwijk, \enquote{A common-path
  interferometer for time-resolved and shot-noise-limited detection of single
  nanoparticles,} {\protect\JournalTitle{Optics Express, Vol. 15, Issue 5, pp.
  2273-2287}} \textbf{15}, 2273--2287 (2007).

\bibitem{Coleal2024}
C.~N. Coleal, J.~W. Wilson, and R.~A. Bartels, \enquote{Transient absorption
  and phase microscopy with balanced detection,}
  {\protect\JournalTitle{https://doi.org/10.1117/12.3028248}} \textbf{13139},
  131390N (2024).

\bibitem{Smith2024}
D.~R. Smith, J.~W. Wilson, S.~Shivkumar, \emph{et~al.}, \enquote{Low-frequency
  coherent raman imaging robust to optical scattering,}
  {\protect\JournalTitle{Chemical and Biomedical Imaging}} \textbf{2}, 584--591
  (2024).

\bibitem{Candeo2019}
A.~Candeo, B.~E. N.~D. Faria, M.~Erreni, \emph{et~al.}, \enquote{A
  hyperspectral microscope based on an ultrastable common-path interferometer,}
  {\protect\JournalTitle{APL Photonics}} \textbf{4} (2019).

\bibitem{Perri2019}
A.~Perri, B.~E.~N. de~Faria, D.~C.~T. Ferreira, \emph{et~al.},
  \enquote{Hyperspectral imaging with a twins birefringent interferometer,}
  {\protect\JournalTitle{Optics Express}} \textbf{27}, 15956 (2019).

\bibitem{dong2019label}
P.-T. Dong, H.~Lin, K.-C. Huang, and J.-X. Cheng, \enquote{Label-free
  quantitation of glycated hemoglobin in single red blood cells by transient
  absorption microscopy and phasor analysis,} {\protect\JournalTitle{Science
  advances}} \textbf{5}, eaav0561 (2019).

\bibitem{born2013principles}
M.~Born and E.~Wolf, \emph{Principles of optics: electromagnetic theory of
  propagation, interference and diffraction of light} (Elsevier, 2013).

\bibitem{Goodman1996}
J.~W. Goodman, \emph{Introduction to Fourier Optics} (McGraw-Hill, 1996), 2nd
  ed.

\bibitem{Mertz2019}
J.~Mertz, \enquote{Introduction to optical microscopy,}
  {\protect\JournalTitle{Introduction to Optical Microscopy}}  (2019).

\bibitem{Hariharan1993}
P.~Hariharan, \enquote{The sénarmont compensator: An early application of the
  geometric phase,} {\protect\JournalTitle{Journal of Modern Optics}}
  \textbf{40}, 2061--2064 (1993).

\bibitem{newlightSellmeier}
N.~Photonics, \enquote{Calcite crystals,}  (2025).
  \url{https://www.newlightphotonics.com/Birefringent-Crystals/Calcite-Crystals}
  [Accessed: 2025-09-20].

\bibitem{Reider2016}
G.~A. Reider, \emph{Photonics An Introduction} (Springer, 2016).

\end{thebibliography}

\end{document}